\documentstyle[aps]{revtex}
\tighten
\draft

\begin{document}
\title{Post-Newtonian Expansion of Gravitational Radiation\footnote{In
{\it Black Holes and Gravitational Waves}, Proceedings of the 9th Yukawa
International Seminar, T. Nakamura and H. Kodama (eds.),
Prog. Theor. Phys. Suppl. No. 136, 146 (1999).}}  
\author{Luc Blanchet} 
\address{D\'epartement d'Astrophysique Relativiste et de
Cosmologie,\\ Centre National de la Recherche Scientifique (UMR
8629),\\ Observatoire de Paris, 92195 Meudon Cedex, France}

\date{September 29, 1999}
\maketitle

\begin{abstract}
The post-Newtonian expansion appears to be a relevant tool for
predicting the gravitational waveforms generated by some astrophysical
systems such as binaries. In particular, inspiralling compact binaries
are well-modelled by a system of two point-particles moving on a
quasi-circular orbit whose decay by emission of gravitational
radiation is described by a post-Newtonian expansion. In this paper we
summarize the basics of the computation by means of a series of
multipole moments of the exterior field generated by an isolated
source in the post-Newtonian approximation. This computation relies on
an ansatz of matching the exterior multipolar field to the inner field
of a slowly-moving source. The formalism can be applied to
point-particles at the price of a further ansatz, that the infinite
self-field of point-particles can be regularized in a certain way. As
it turns out, the concept of point-particle requires a precise
definition in high post-Newtonian approximations of general
relativity.
\end{abstract}

\section{Introduction}

Binary systems of compact objects (neutron stars or black holes)
emitting gravitational waves in their late stage of evolution leading
to a final coalescence might play the major role in the XXIth-century
gravitational-wave astronomy. In the final stage, the two objects
collide; for instance two black holes form a single black hole which
rings down and emits gravitational waves before settling down into a
stationary configuration. We refer to the papers in this volume
devoted to numerical relativity \cite{num} and to the close-limit
approximation \cite{close} for descriptions of the coalescence
phase.
 
In the earlier phase, preceding immediately the coalescence, the two
objects undergo a long (adiabatic) inspiral driven by the emission of
gravitational radiation, or equivalently by the radiation reaction
forces applied to the orbit. In this paper we are mainly interested
in the inspiral phase of compact binaries. During this phase the
gravitational radiation is essentially produced by the dynamical
motion of the two compact objects. In principle, the dynamics can be
well-approximated by a post-Newtonian expansion of general relativity.
Given the highly relativistic nature of inspiralling compact binaries
(the orbital velocity can reach 30\% of the speed of light in the
last rotations), the problem is just that of pushing the
post-Newtonian approximation farther enough in order to be useful to
future observations. In recent years it has been realized by several
groups
\cite{3mn,CFPS93,FCh93,CF94,TNaka94,BSat95,P95,DIS98}
that in the case of inspiralling neutron star binaries the
post-Newtonian expansion should be controlled up to the very high
3PN order. During the inspiral, the internal
structure of the stars plays a little role, and one can conveniently
describe the two compact bodies by ``point-particles" (what only
matters are the two masses, but apart from that the objects can be
ordinary neutron stars, or black holes or even naked singularities).
Note that a priori the concept of point-particle does not make sense
in general relativity except in the test-mass limit. However within a
post-Newtonian approximation one can give a sense to what we call a
point-particle; but even there, particularly when going to high
post-Newtonian orders, the concept of a point-particle is non-trivial
and must be carefully defined.

A post-Newtonian computation of the inspiral of compact binaries can
be based on the following strategy. One starts by implementing a
general formalism for the dynamics and the gravitational-wave emission
of a slowly-moving isolated source. By slowly-moving we mean the
existence of a small post-Newtonian parameter, say

\begin{equation}
\varepsilon={\rm Max}~\biggl\{ \biggl|{T^{0i}\over T^{00}}\biggr|, 
\biggl|{T^{ij}\over T^{00}}\biggr|^{1/2}\biggr\} \ ,
\end{equation}
where $T^{\mu\nu}$ denotes the matter stress-energy tensor in some
Cartesian coordinate system covering the source;\footnote{We assume
that the source is self-gravitating, so that $\varepsilon\sim
(GM/ac^2)^{1/2}$ where $M$ and $a$ denote the mass and radius.}  Greek
indices take values 0,1,2,3, and Latin 1,2,3. We allow $\varepsilon$
to be as large as $30\%$ in order to cover the case of inspiralling
compact binaries. As we see, strictly speaking the source is
``slowly-moving", but in fact it can be quite relativistic. As usual,
neglecting terms in the gravitational field according to their formal
order in $\varepsilon$ is valid in the so-called near-zone of the
source; the region for which $r/\lambda=O(\varepsilon)$, where
$\lambda$ is a typical wavelength of the radiation field.

The crucial demand on any wave-generation formalism is to be able to
relate the gravitational waveform far from the source (to order $1/R$
in the distance to the source at retarded times) to the matter
stress-energy tensor $T^{\mu\nu}$. This problem of relating the
retarded far-field to the source's matter content is extremely
difficult within the exact theory, because of the non-linearities of
the field equations; however, the solution exists in a framework of
post-Newtonian approximations for slowly-moving sources. Another
requirement for a general formalism is for its ability to control the
equations of motion of the source, and in particular the gravitational
radiation reaction forces therein.  The radiation reaction forces are
to be consistent with the radiation field at infinity (by definition),
and thus to depend on boundary conditions such as the no-incoming
radiation condition imposed at past null infinity (which ensures that
the source is physically isolated). In a post-Newtonian approximation,
a difficulty of the problem is that the radiation reaction forces
enter the equations of motion of the source which are determined by
the post-Newtonian expansion valid only in the near zone. Therefore
one must supplement the post-Newtonian expansion by a condition of
matching the near-zone field to the radiation field [see (2$\cdot$5)
below].  In this article we outline a particular post-Newtonian
formalism, making extensive use of multipole moments, which is issued
from work of Blanchet and Damour, \cite{BD86,BD89,BD92} Damour and
Iyer, \cite{DI91} and Blanchet. \cite{B95,B98tail,B98mult} A different
formalism recently defined by Will and Wiseman \cite{WWi96} on
foundations laid by Epstein and Wagoner \cite{EW75} and Thorne
\cite{Th80} is described by C. Will in this volume.

Note that the present post-Newtonian formalism is interested in the
{\it formal} post-Newtonian expansion $\varepsilon\to 0$; in
particular it does not try to investigate the exact mathematical
nature of the post-Newtonian series. Thus the connection between the
approximation and the exact theory is not controlled.\footnote{A mean
to understand the limit relation of Einstein's theory to Newton's is
to introduce a frame theory {\it \`a la} Ehlers.  \cite{Ehlers}}
Simply, one looks for a mathematically well-defined scheme for
generating the successive post-Newtonian approximations. The hope is
that in applications, when comparing to observations the
post-Newtonian prediction developed to high enough order in
$\varepsilon$, we shall drastically limit the magnitude of systematic
errors, resulting in a very accurate measurement.

Be aware that at present no systematic algorithm for generating the
post-Newtonian series of the near-zone gravitational field exists. One
can express the multipole moments of the source in terms of the
infinite formal post-Newtonian series, \cite{B98mult} but one does not
know how to compute {\it explicitly} this series to all orders, in the
manner, say, of the explicit algorithm for generating the post-{\it
Minkowskian} series in the source's exterior. \cite{BD86} But of
course, the first few post-Newtonian approximations, up to, say, the
2.5PN order, are very well understood.
\cite{Fock,ChandraN,ChandraE,Ehlers80,Kerlick80,PapaL} A difficulty
associated with the post-Newtonian iteration is the appearance in
higher post-Newtonian orders of Poisson-type integrals which are
divergent at the bound at infinity. This comes from the fact that the
post-Newtonian iteration is valid only in the near-zone, and so the
post-Newtonian coefficients rapidly blow up far from the system. This
signals that the standard Poisson integral does not constitute the
relevant solution of the Poisson equation in this context.

A crucial element of the present formalism is that it is {\it a
  priori} only valid for continuous (in fact, smooth) matter
  distributions. Thus, $T^{\mu\nu}$ is assumed from the start to be
  regular, for instance to describe a smooth hydrodynamical
  fluid. This excludes {\it a priori} the very interesting application
  to ``self-gravitating" point-particles (as opposed to ``test"
  point-particles), whose self-field becomes infinite at the location
  of the particle, thus creating a singularity. So we are obliged to
  introduce a new ingredient. Our proposal is that this be a
  regularization {\it \`a la} Hadamard\cite{Hadamard,Schwartz} for
  removing systematically the infinite self-field of the
  particles. Only when we assume a regularization, can we use a
  $T^{\mu\nu}$ constituted of Dirac delta-functions. The removing of
  the divergent terms is for the moment done without further
  justification; simply this is an ansatz which, as far as we can see,
  yields consistent computations in practice, and which has been
  checked to yield the correct result in some cases, but which remains
  an ansatz. Notably we do not prove that the regularization is still
  permissible in higher post-Newtonian approximations, or even that it
  is possible to find a consistent regularization at all, or that the
  result of two different (though consistent) regularizations would be
  the same. We will content ourselves with the use of the Hadamard
  regularization which yields in practice, to the first few
  post-Newtonian orders, some consistent (and rather elegant)
  computations.

A different strategy is possible when we have at our disposal a
natural background space-time. This is the case when the mass ratio of
two particles is so small that one can view one of them as moving in
the Schwarzschild or Kerr background generated by the other. In the
small mass-ratio limit the particle moves on a geodesic of the
background, and we can compute the emitted radiation using a linear
background perturbation. This approach has reached a mature state:
notably the gravitational radiation emitted by a test particle in
orbit around a Schwarzschild black hole was computed to very high
post-Newtonian order;\cite{Sasa94,TSasa94,TTS96} this computation
represents an important benchmark against which the standard
post-Newtonian expansion can be checked. To second-order in the
black-hole perturbation there are terms vanishing in the mass-ratio
limit, the particle reacts to the gravitational radiation emission,
and again we face the problem of how to regularize the particle's
self-field.

\section{General sources}

For a general compact-support stress-energy tensor $T^{\mu\nu}$, we
want to solve the field equations of general relativity in the form of
a post-Newtonian expansion $\varepsilon\to 0$. We reduce the field
equations by means of the condition of harmonic coordinates, that is
$\partial_\nu h^{\mu\nu} = 0$ where $h^{\mu\nu}\equiv \sqrt{-g}
g^{\mu\nu}-\eta^{\mu\nu}$, so that

\begin{equation}
\Box h^{\mu\nu} = {16\pi G\over c^4}|g| T^{\mu\nu} 
+ \Lambda^{\mu\nu}(h,\partial h,\partial^2 h) \ ,
\end{equation} 
where we have introduced a Minkowskian background, $\eta^{\mu\nu}={\rm
diag}(-1,1,1,1)=\eta_{\mu\nu}$ (with respect to which all indices are
raised and lowered), and the associated d'Alembertian operator,
$\Box\equiv\Box_\eta=\eta^{\mu\nu}\partial_\mu\partial_\nu$. The
gravitational source term $\Lambda^{\mu\nu}$ is a complicated
functional of the field and its first and second space-time
derivatives. We define the total stress-energy (pseudo-)tensor
$\tau^{\mu\nu}$ of the matter and gravitational fields as

\begin{equation}
\tau^{\mu\nu}=|g| T^{\mu\nu} + {c^4\over 16\pi G}\Lambda^{\mu\nu} \ .
\end{equation}
Of course $\tau^{\mu\nu}$ is not a generally-covariant tensor, but it
is a Lorentz tensor relatively to our Minkowskian background. It is
conserved in a Lorentz-covariant sense, and this is equivalent to the
covariant conservation of $T^{\mu\nu}$,

\begin{equation}
\partial_\nu\tau^{\mu\nu}=0\quad\Leftrightarrow\quad\nabla_\nu
T^{\mu\nu} =0 \ .
\end{equation}
The propagation of $h^{\mu\nu}$ subject to the Einstein field
equations (2$\cdot$1) is a well-posed problem, for which we need to
choose some initial conditions in the past. In this paper we shall
assume that the field is stationary before some remote date $-{\cal
T}$, so that there is no radiation generated by sources at infinity
incoming onto the system. Arguably, the condition of stationarity in
the past is too strong; for instance it does not cover a physical
situation where two bodies moving initially on unbound
(hyperbolic-like) orbits would form a bound system. However one may in
some cases justify the assumption {\it a posteriori}, by checking that
the formulas obtained under it are still valid in a more general
physical situation such as the initial scattering of two bodies. With
no-incoming radiation one can transform the differential Einstein
equations (2$\cdot$1) into the integro-differential equations

\begin{equation}
 h^{\mu\nu} = {16\pi G\over c^4} \Box ^{-1}_{R}
\tau^{\mu\nu}\equiv-{4G\over c^4}\int {d^3{\mbox{\boldmath$x$}}'\over
|{\mbox{\boldmath$x$}}-{\mbox{\boldmath$x$}}'|}\tau^{\mu\nu}
\left({\mbox{\boldmath$x$}}',t-|{\mbox{\boldmath$x$}}
-{\mbox{\boldmath$x$}}'|/c\right)
\ ,
\end{equation}
where $\Box^{-1}_{R}$ denotes the standard retarded inverse
d'Alembertian.

In our approach we resolve the wave-generation problem by finding the
solution of the field equations (2$\cdot$4) in the {\it exterior}
region of the source (outside the compact support of $T^{\mu\nu}$) by
means of an infinite multipole-moment series for $h^{\mu\nu}$, that we
denote ${\cal M}(h^{\mu\nu})$. This means in particular that we must
be able to relate the multipole moments parametrizing this series to
the matter content of the source. In general this is not an easy task,
but this can be done using the post-Newtonian expansion, in the
physical case where the source is slowly-moving. The latter source
multipole moments are then ``propagated" with the help of a
post-Minkowskian expansion to large distances from the source, and
related there to the so-called radiative multipole moments directly
accessible to a far-away observer.\footnote{Using a post-Minkowskian
expansion for the exterior field in conjunction with the multipolar
series is an old idea of Bonnor,\cite{Bo59} later generalized by
Thorne.\cite{Th80}}
 
To obtain the expression of the multipole moments of the source in
terms of $T^{\mu\nu}$ we use an asymptotic matching between the
multipole expansion ${\cal M}(h^{\mu\nu})$, valid everywhere outside
the source, and the post-Newtonian expansion denoted ${\overline
h}^{\mu\nu}$, valid in the near-zone. For slowly-moving sources the
two domains of validity of the multipole and post-Newtonian expansions
overlap in the so-called exterior near-zone, and we can write there
the numerical equality ${\cal M}(h^{\mu\nu})={\overline h}^{\mu\nu}$.
Then we transform this equality into a ``matching equation", that is
an equation between two series of the same nature, which is formally
valid ``everywhere". For this purpose we replace the multipole terms
on the left-hand side of the equality by their formal post-Newtonian
expansions (this means expanding all retardations $t-r/c$ when
$c\to\infty$; so in fact the latter expansion is equivalent to a
near-zone expansion $r\to 0$), and we consider on the right-hand side
the multipole decomposition of (each of the coefficients of) the
near-zone expansion. Therefore the matching equation reads

\begin{equation}
\overline {{\cal M}(h^{\mu\nu})}  = {\cal M}({\overline h}^{\mu\nu}) \ .  
\end{equation} 
Satisfying this equation permits in principle to find the unique
physical solution of the field equations valid inside and outside the
source (and, in the present case, in the uniquely defined harmonic
coordinate system).

Let us simply announce here the result\cite{B95,B98mult} for the
expression of ${\cal M}(h^{\mu\nu})$ satisfying at once the vacuum
Einstein field equations (with no-incoming radiation) and the matching
equation (2$\cdot$5). There are two terms,

\begin{equation}
{\cal M}(h^{\mu\nu}) = - {4G\over c^4} \sum^{+\infty}_{l=0}
{(-)^l\over l!} \partial_L \left\{ {1\over r} {\cal F}^{\mu\nu}_L
(t-r/c) \right\}+{\rm FP}_{B=0}\, \Box^{-1}_R [ r^B {\cal
M}(\Lambda^{\mu\nu})] \ .
\end{equation} 
The second term constitutes a {\it particular} solution of the vacuum
field equations, which is defined by analytic continuation in a
complex parameter $B$ as being the finite part of the Laurent
expansion when $B\to 0$ (in our notation ${\rm FP}_{B=0}$). The reason
for the need of a finite part is that the integrand of the retarded
integral is in the form of a multipole expansion which is not valid
inside the source, and which is actually singular at the spatial
origin of the coordinates $r=0$ located inside the source. Now the
first term in (2$\cdot$6) (in which we denote $L=i_1\cdots i_l$ and
$\partial_L\equiv \partial_{i_1}\cdots\partial_{i_l}$) represents
clearly an {\it homogeneous} solution of the wave equation. This
solution is uniquely singled out so that when added to the particular
solution it ensures the satisfaction of the matching equation
(2$\cdot$5). The ``multipole moments" parametrizing this solution are
given as\cite{B98mult}

\begin{equation}
{\cal F}^{\mu\nu}_L (u) = {\rm FP}_{B=0} \int d^3
 {\mbox{\boldmath$x$}}~| {\mbox{\boldmath$x$} }|^B {\hat x}_L
 \int^1_{-1} dz~ \delta_l(z) {\overline \tau^{\mu\nu}}
 ({\mbox{\boldmath$x$}}, u+z|{\mbox{\boldmath$x$}}|/c) ,
\end{equation}
(where $u=t-r/c$).  The moments are also defined by analytic
continuation in $B$, which makes them well-defined mathematically
(note that the integrand behaves typically as a positive power of the
distance at infinity, and thus that the integral would be strongly
divergent at infinity without any finite part). In (2$\cdot$7) we use
a special notation for a symmetric trace-free product of vectors:
${\hat x}_L \equiv {\rm STF}(x_L)$ where $x_L\equiv x_{i_1}\cdots
x_{i_l}$ (we have $\delta_{i_li_{l-1}}{\hat x}_L=0$). The
$z$-integration in (2$\cdot$7) involves the weighting function

\begin{equation}
\delta_l (z) = {(2l+1)!!\over 2^{l+1} l!} (1-z^2)^l
\ ; \quad\int^1_{-1} dz~\delta_l (z) = 1 \ .
\end{equation} 
In the limit of large $l$ we have $\lim_{l\to\infty}\delta_l=\delta$
the Dirac measure.

The crucial point about the multipole moments (2$\cdot$7) is that they
are generated by the {\it post-Newtonian} expansion of the
stress-energy pseudo-tensor, that is ${\overline \tau}^{\mu\nu}$,
rather than $\tau^{\mu\nu}$ itself. It is at this point that our
assumption of matching with a slowly-moving post-Newtonian source
enters.  Note that although the integrand of the multipole moments is
in the form of a post-Newtonian expansion valid only in the near-zone,
the integration is to be performed on the whole 3-dimensional space
(see Ref. \cite{B98mult} for details). This is one of the beauties of
the analytic continuation, that it permits to handle integrals over
$I\!\!R^3$ without introducing a cutoff at the edge of the near-zone.
On the contrary, Will and Wiseman\cite{WWi96} do not use analytic
continuation, and consider integrals extending only over the
near-zone. It can be shown that the present formalism is equivalent to
that of Will and Wiseman.

With the expression (2$\cdot$7) in hands, it is straightforward to
define six sets of irreducible multipole moments $\{I_L, J_L, W_L,
X_L, Y_L, Z_L\}$ associated with the six independent components of
${\cal F}^{\mu\nu}_L$ (ten minus four because of the harmonic
coordinate condition). This constitutes our primary definition for the
source multipole moments. With more work (considering a coordinate
transformation in the exterior zone) we can further define only two
sets of irreducible moments $\{M_L, S_L\}$ which are less directly
connected to the source but are very useful in practical computations.

Convenient notions of the source multipole moments being chosen, let
us relate them to the radiative moments at infinity. We follow the
standard definition\cite{Th80} that the radiative moments parametrize
in radiative (Bondi-type) coordinates the leading term $1/R$ in the
distance to the source. Clearly, looking at (2$\cdot$6), we see that
the radiative moments have two contributions. Essentially the
contribution coming from the first term of (2$\cdot$6) is that of the
source multipole moments $I_L, J_L,\cdots, Z_L$. In a linear theory
the radiative moments would contain only this contribution, i.e., they
would agree with the source moments.  But in general relativity we
have also the contribution of the second term in (2$\cdot$6) (which is
at least quadratic in the field strength $G$), so there will be many
non-linear interactions between the source moments to any order in
$G$. Therefore the radiative moments are given by some (very
complicated) non-linear functionals of the source moments. We compute
these with the help of a post-Minkowskian algorithm.\cite{BD86,B87}
That is, we are able to rewrite (2$\cdot$6) in the form of a formal
expansion when $G\to 0$,

\begin{equation}
 {\cal M} (h^{\mu\nu}) = Gh^{\mu\nu}_1[I_L,J_L,\cdots] 
+ G^2h^{\mu\nu}_2+\cdots + G^n h^{\mu\nu}_n+\cdots \ ,
\end{equation}
where the first term is a solution of the linearized field equations
which depends on the source moments [this is essentially the first
term in (2$\cdot$6)], and where all the subsequent non-linear
corrections are constructed by post-Minkowskian iteration (see
Ref. \cite{BD86} for the proof that this can be done to any
post-Minkowskian order). Next we change coordinates from harmonic to
radiative ones $(T,{\mbox{\boldmath$X$}})$ and expand the metric when
$R\to\infty$. All the physical information about the radiation field
is contained into the so-called transverse-traceless (TT) projection
of the spatial ($ij$) metric. The radiative moments $\{U_L, V_L\}$ are
defined from the term of order $1/R$ as

\begin{eqnarray}
 {\cal M}(h_{ij})^{\rm TT} ({\mbox{\boldmath$X$}} ,T)&=&-{4G\over
 c^2R} {\cal P}_{ijab} \sum_{l \geq 2} {1\over c^l l !} \left\{
 N_{L-2} U_{abL-2} - {2l \over c(l+1)} N_{cL-2} \varepsilon_{cd(a}
 V_{b)dL-2}\right\}\nonumber\\ &&+O\left( {1\over R^2}\right) \ ,
\end{eqnarray}
where $N_i=X^i/R$, $N_{L-2}=N_{i_1}\cdots N_{i_{l-2}}$, etc., and
where the TT projector reads ${\cal P}_{ijab}= (\delta_{ia}
-N_iN_a)(\delta_{jb} -N_jN_b) -{1\over 2} ( \delta_{ij} -N_iN_j)
(\delta_{ab}-N_aN_b)$. Now it remains to compare the result of the
iteration of the field (2$\cdot$9) with the relation (2$\cdot$10) in
order to deduce the radiative moments $U_L$ and $V_L$ in terms of the
source moments, and therefore in terms of $T^{\mu\nu}$ via the
post-Newtonian expansion of the pseudo-tensor $\tau^{\mu\nu}$ [see
(2$\cdot$7)].

The wave-generation is complete in this way. Note that the formalism
is general in the sense that we did not specify a particular form of
$T^{\mu\nu}$. Thus, for application to a particular problem, we need
first to choose a model of $T^{\mu\nu}$, for instance a perfect fluid
or a model of point-particles (see below), and insert that model into
the formalism.

\section{Point-particles}

The choice of a model of point-particles in (post-Newtonian) general
relativity is non-trivial because it is intimately related with the
choice of a regularization for removing the self-field of
point-particles. Of course, for a {\it test} particle moving on a
fixed smooth gravitational background $g_{\mu\nu}^{ B}$, there is no
problem and the stress-energy tensor reads

\begin{eqnarray}
T^{\mu\nu}_{\rm test}({\mbox{\boldmath$x$}},t)={m v^\mu v^\nu \over
\sqrt{-g_{\rho\sigma}^{B}(y) v^\rho v^\sigma/c^2}} {\delta
({\mbox{\boldmath$x$}}-{\mbox{\boldmath$y$}}(t))\over
\sqrt{-g^{B}(x)}} \ ,
\end{eqnarray}
where $v^\mu=dy^\mu/dt$ is the particle's coordinate velocity and
$\delta$ is the 3-dimensional Dirac measure (a particular case of
distribution).

However, in the case of ``{\it self-gravitating}" particles
contributing to the gravitational field, the previous stress-energy
tensor does not make sense. Indeed, already at the Newtonian order,
the metric $g_{\mu\nu}$ generated by two particles for instance
contains the Newtonian potential $U=G m_1/r_1+G m_2/r_2$ where
$r_{1,2}=|{\mbox{\boldmath$x$}}- {\mbox{\boldmath$y$}}_{1,2}|$ is the
distance between the field point and the source points. Clearly $U$ is
infinite at the location of the particles, and must be regularized in
some way in order to speak of its value at point 1, say $(U)_1$. We
know that the correct result in Newtonian gravity is $(U)_1=G
m_2/r_{12}$ where
$r_{12}=|{\mbox{\boldmath$y$}}_1-{\mbox{\boldmath$y$}}_2|$. But in
high post-Newtonian orders we shall meet more difficult quantities
such as $(U^4)_1$ which necessitate a precise definition.

Let us model the stress-energy tensor of two point-particles in
post-Newtonian approximations of general relativity by

\begin{eqnarray}
T^{\mu\nu}( {\mbox{\boldmath$x$} },t)= {m_1 v_1^\mu v_1^\nu \over
\sqrt{-(g_{\rho\sigma})_1 v_1^\rho v_1^\sigma/c^2}} {\delta (
{\mbox{\boldmath$x$} }- {\mbox{\boldmath$y$}_1}(t))\over
\sqrt{-g(x)}}+1\leftrightarrow 2 \ ,
\end{eqnarray}
where $1\leftrightarrow 2$ means the exchange of 1 to 2 on the two
previous terms and, $(g_{\rho\sigma})_1$ denotes the {\it regularized}
value of the metric $g_{\rho\sigma}$ at the location of the particle
1, according to a procedure based on the
Hadamard\cite{Hadamard,Schwartz} partie finie of a singular function
and a divergent integral. Note that in fact the chosen regularization
{\it defines} our model of point-particles. A conjecture would be that
another type of regularization, if sufficiently powerful to produce
unambiguous results at high post-Newtonian orders (as seems to be the
case of the Hadamard regularization), would yield identical physical
results.

To define the Hadamard regularization we introduce an appropriate
class of functions ${\cal F}$ on $I\!\!R^3$. By definition
$F({\mbox{\boldmath$x$} })$ belongs to ${\cal F}$ if and only if it is
smooth on $I\!\!R^3$ except at two singular points
${\mbox{\boldmath$y$}}_1$ and ${\mbox{\boldmath$y$}}_2$, around which
it admits a power-like expansion of the type

\begin{equation}
F=\sum_a r_1^a f_{1(a)}({\mbox{\boldmath$n$}}_1) \qquad \hbox{when
$r_1\to 0$}
\end{equation}
(and idem for 2). The summation index $a$ is assumed to be bounded
from below, $a\geq -a_0$ (no essential singularity), and to take real
discrete values, say $a\in \{a_j\}_{j\in I\!\!N}$ where $a_j\in
I\!\!R$. In fact in most cases one can assume $a\in Z\!\!\!Z$. As
indicated the coefficients $f_{1(a)}$ depend on the unit direction
${\mbox{\boldmath$n$}}_1$ of approach to the singularity.  In
(3$\cdot$3) we do not write any remainder for the expansion because we
shall need only formulas depending on the $f_{1(a)}$, and never the
complete expansion itself. \footnote{In addition to (3$\cdot$3) we
assume that the functions $F\in {\cal F}$ decrease sufficiently
rapidly when $|{\mbox{\boldmath$x$}}|\to\infty$, so that all integrals
we consider are convergent at infinity.}

First we define the Hadamard partie finie of the singular function $F$
at the location of the singularity 1 for instance, say $(F)_1$. Simply
we pick up the coefficient of the zeroth power of $r_1$ in
(3$\cdot$3), namely $f_{1(0)}$, and average over all directions
${\mbox{\boldmath$n$}}_1$:

\begin{equation}
(F)_1\equiv \int {d\Omega_1\over 4\pi} f_{1(0)} \ .
\end{equation}
Second we define the partie finie (in short ${\rm Pf}$) 
of the divergent integral $\int d^3{\mbox{\boldmath$x$}}~F$. We remove
from $I\!\!R^3$ two spherical balls surrounding the two singularities,
of the form $r_{1,2}\leq s$, where $s$ is a small radius. Using (3$\cdot$3)
it is easy to determine the expansion when $s\to 0$ of the integral
extending on $I\!\!R^3$ deprived from these two balls, next to
subtract all the terms which are divergent when $s=0$, and then to take
the limit $s\to 0$ of what remains. The result is the Hadamard partie
finie. Actually we find some logarithms of $s$ in the expansion, so
there is an ambiguity linked with the freedom of choosing a unit of
length to measure $s$. To account with this ambiguity we introduce a
constant length scale $s_1$ (and similarly $s_2$). We find

\begin{eqnarray}
{\rm Pf}\int d^3 {\mbox{\boldmath$x$}}~ F&\equiv&\lim_{s\to
0}\biggl\{\int_{r_1>s\atop r_2>s} d^3{\mbox{\boldmath$x$}}~F
+\!\!\!\sum_{a+3\leq -1}{s^{a+3}\over a+3}\int d\Omega_1
f_{1(a)}\nonumber\\ &&\quad\quad+\ln\left({s\over s_1}\right)\int
d\Omega_1 f_{1(-3)} +1\leftrightarrow 2\biggr\} \ ,
\end{eqnarray}

The two definitions (3$\cdot$4) and (3$\cdot$5) are closely related to
each other.  To see this, apply (3$\cdot$5) to the case where the
function is actually a gradient $\partial_iF$ (it is clear that $F\in
{\cal F}$ implies $\partial_iF\in {\cal F}$). We find

\begin{equation}
{\rm Pf}\int d^3{\mbox{\boldmath$x$} }~ \partial_iF=-4\pi 
(n_1^i r_1^2F)_1-4\pi (n_2^i r_2^2F)_2 \ ,
\end{equation}
whose proof involves the Gauss theorem on the two surfaces $r_{1,2}=s$
surrounding the two singularities. Thus, for singular functions, the
integral of a gradient is not zero. This indicates that the
``ordinary" derivative is not adequate for the purpose of application
of a ``fluid" formalism to point-particles (because the integral of a
gradient is always zero for fluids). Instead one must generalize the
derivative to take into account the singularities, in a way similar to
the distributional derivative of distribution theory.\cite{Schwartz}

Another nice connection between the two definitions (3$\cdot$4) and
(3$\cdot$5) is

\begin{equation}
\lim_{\epsilon\to 0}{\rm Pf}\int d^3{\mbox{\boldmath$x$} }~
\delta_\epsilon ({\mbox{\boldmath$x$} }-{\mbox{\boldmath$y$} }_1)
F({\mbox{\boldmath$x$} })=(F)_1 \ ,
\end{equation}
where $\delta_\epsilon$ denotes the Riesz\cite{Riesz}
delta-function defined for any $\epsilon>0$ by

\begin{equation}
\delta_\epsilon ({\mbox{\boldmath$x$} })={\epsilon (1-\epsilon)\over
4\pi} |{\mbox{\boldmath$x$} }|^{\epsilon-3} \ , \qquad\mbox{so
that}\qquad\Delta |{\mbox{\boldmath$x$} }|^{\epsilon-1}=-4\pi
\delta_\epsilon \ .
\end{equation}
Clearly, in the limit $\epsilon\to 0$, the Riesz delta-function yields
a generalization of the Dirac measure applicable to the Hadamard
partie finie of a singular function in ${\cal F}$.

To conclude, let us give the result of the application of the general
formalism to a binary system of point-particles, concerning the total
flux (or ``gravitational luminosity" ${\cal L}$) emitted by the binary
in the form of gravitational waves. This quantity plays a crucial role
in the computation of the orbital phase of the binary, as it evolves
with time taking into account the loss of energy by gravitational
waves. For circular orbits the orbital phase $\phi=\int \omega dt$ is
obtained from the energy balance equation as

\begin{equation}
{dE\over dt}=-{\cal L}\qquad\Rightarrow\qquad \phi=-\int {\omega
dE\over {\cal L}} \ ,
\end{equation}
where $E$ denotes the orbital binding energy of the binary in the
center of mass frame.  Let $m_1$ and $m_2$ be the two masses, and
denote $m=m_1+m_2$ and the mass ratio $\nu=\mu/m=m_1 m_2/m^2$. As a
small post-Newtonian parameter we define $x=(Gm\omega/c^3)^{2/3}$
where $\omega=2\pi/P$ is the orbital frequency and $P$ the period [$x$
is of order $O(\varepsilon^2)$ in the notation (1$\cdot$1)]. The
present post-Newtonian accuracy for ${\cal L}$ is 3.5PN except that
the contributions proportional to the mass ratio $\nu$ in the 3PN term
are not yet under control. These contributions are indicated by
$O(\nu)$ in the formula below; their computation is a work in progress
(collaboration with Faye, Iyer and Joguet). We obtain

\begin{eqnarray}
{\cal L} = {32c^5\over 5G} \nu^2 x^5 \biggl\{1 &+&
\left(-\frac{1247}{336}-\frac{35}{12}\nu \right) x + 4\pi
x^{3/2}\nonumber\\ &+&\left(-\frac{44711}{9072}+\frac{9271}{504}\nu
+ \frac{65}{18} \nu^2\right) x^2 \nonumber \\
&+&\left(-\frac{8191}{672}-\frac{535}{24}\nu \right)\pi x^{5/2}
\nonumber \\ &+&\left(
\frac{6643739519}{69854400}-\frac{1712}{105}C-\frac{856}{105}
\ln(16x)+\frac{16}{3}\pi^2+O(\nu) \right)x^3 \nonumber \\
&+&\left(-\frac{16285}{504}+\frac{176419}{1512}\nu
+\frac{19897}{378}\nu^2 \right)\pi x^{7/2}+O(x^4)\biggr\} \ .
\end{eqnarray}
The Newtonian approximation was given by Landau and Lifschitz\cite{LL}
and the classic work of Peters and Mathews\cite{PM} (recall that the
Newtonian term is sufficient to account for the radiation emitted by
the binary pulsar PSR1913+16).  The 1PN term was computed long ago by
Wagoner and Will\cite{WagW76} applying the formalism of Epstein and
Wagoner,\cite{EW75} and later confirmed\cite{BS89} applying the
present formalism. The 1.5PN term (with $4\pi$ as a factor) is due to
gravitational-wave tails, which was first obtained by
Poisson\cite{P93} using a black-hole perturbation, and it was
re-computed by Wiseman\cite{Wi93} and Blanchet and
Sch\"afer\cite{BS93} within the present formalism. The 2PN term was
computed independently by Blanchet, Damour and Iyer\cite{BDI95}
applying the same formalism, and by Will and Wiseman\cite{WWi96}
applying their new formalism (see Ref. \cite{BDIWW95} for a
summary). The 2.5PN and 3.5PN orders are due to higher-order tail
effects.\cite{B96,B98tail} The 3PN order is very interesting. It
involves a logarithmic term (first obtained in this context by Tagoshi
and Nakamura,\cite{TNaka94}) as well as the Euler constant
$C=0.577\cdots$ and a $\pi^2$. These contributions in the 3PN order,
computed in Ref. \cite{B98tail}, are due to the so-called tails of
tails, that is to say, tails generated by the tails themselves --- a
purely cubic effect.  The corrections $O(\nu)$ in the 3PN term depend
on the full details concerning the regularization outlined previously.

Using $x$ as a post-Newtonian parameter is convenient because of its
invariant meaning (the same in all coordinate systems). Thus the
formula (3$\cdot$10) can be directly compared with the one obtained
using a black-hole perturbation (which typically describes the
background in Schwarzschild coordinates) triggered by the motion of a
test particle of mass $\nu m$ in the field of a black-hole of mass
$m$. Using this method Sasaki, Tagoshi and
Tanaka\cite{Sasa94,TSasa94,TTS96} obtained ${\cal L}$ to 5.5PN order
in the limit $\nu\to 0$. In the same limit the formula (3$\cdot$10)
agrees completely with these results to the corresponding
order. Notably the rational fraction 6643739519/69854400 in the 3PN
coefficient comes out exactly the same in the post-Newtonian formalism
as in the black-hole perturbation approach (collaboration with Iyer
and Joguet).

\section*{Acknowledgements}

I wish to thank the organizers, especially T. Nakamura, for inviting
me to this interesting and successful conference.

\end{document}